\documentclass{kluwer}    

\newdisplay{guess}{Conjecture}

\usepackage{graphicx}

\begin{document}                                                                                   
\begin{article}
\begin{opening}         
\title{The ionized  gas in the Galactic Center Radio Arc}
\author{N.\, J. \surname{Rodr\'{\i}guez-Fern\'andez}\thanks{Partially
supported by {\em Consejer\'{\i}a de Educaci\'on de la Comunidad de Madrid}}} 
\author{J. \surname{Mart\'{\i}n-Pintado}}
\author{P. \surname{de Vicente}}             
\runningauthor{Rodr\'{\i}guez-Fern\'andez et al.}
\runningtitle{Ionized        gas in the Galactic Center}
\institute{Observatorio Astron\'omico Nacional,
Apdo. 1143, E28800 Alcal\'a de Henares, Spain}
\date{October 15, 2001}


\end{opening}           

The Radio Arc is one of the most prominent
radio continuum features in the Galactic center region.
It is composed by long and thin filaments that emit non-thermal radiation
(Non-Thermal Filaments; NTFs)
and indicate the presence of a strong component of the magnetic field
perpendicular to the Galactic plane.
The Arc is apparently connected to Sgr~A by a ``bridge'' 
of arched filaments that emit
thermal radio continuum (Thermal Filaments).
There are two other thermal sources in the vicinity of the NTFs: 
the Sickle and the Pistol Nebula.

The origin of the ionization of those thermal  features has been a subject
of great interest in the last years. It was first thought
that they were the surfaces of molecular clouds ionized by collisions
with the relativistic particles that illuminate the NTFs.
However, with the discovery of the outstanding clusters of young stars
known as the Quintuplet and the Arches clusters, the effect of UV radiation
on the ionization of the thermal features has been revised.
There is increasing evidence that the Sickle and the Pistol Nebula are
ionized by the Quintuplet and that the Arches cluster could
account for the ionization of the Thermal Filaments
(see  Rodr\'{\i}guez-Fern\'andez et al. 2001, A\&A 377, 631 ---hereafter
RF01--- and references therein).

\begin{figure}[tbhp!]
\centerline{
\includegraphics[width=5cm]{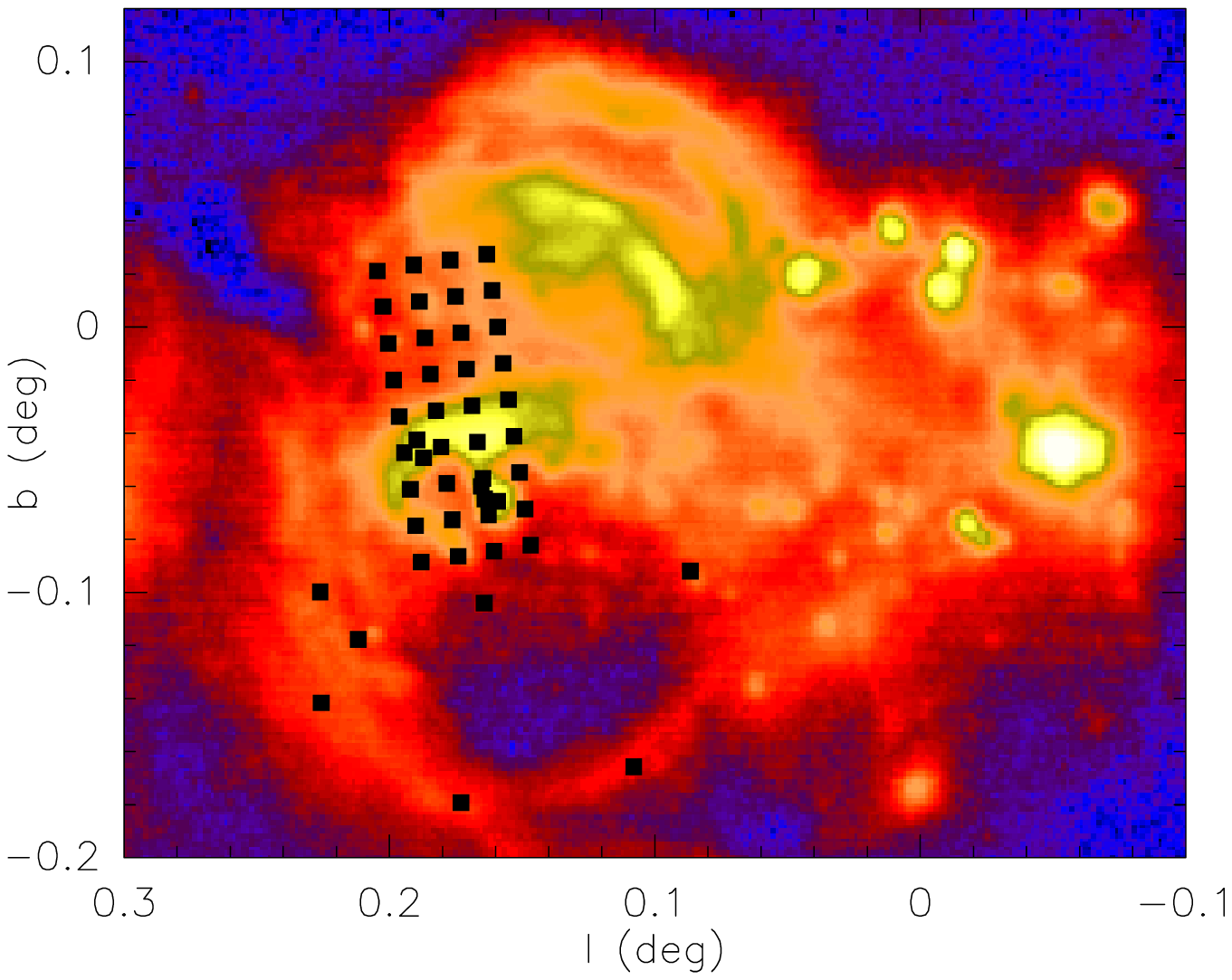}
\includegraphics[width=5cm]{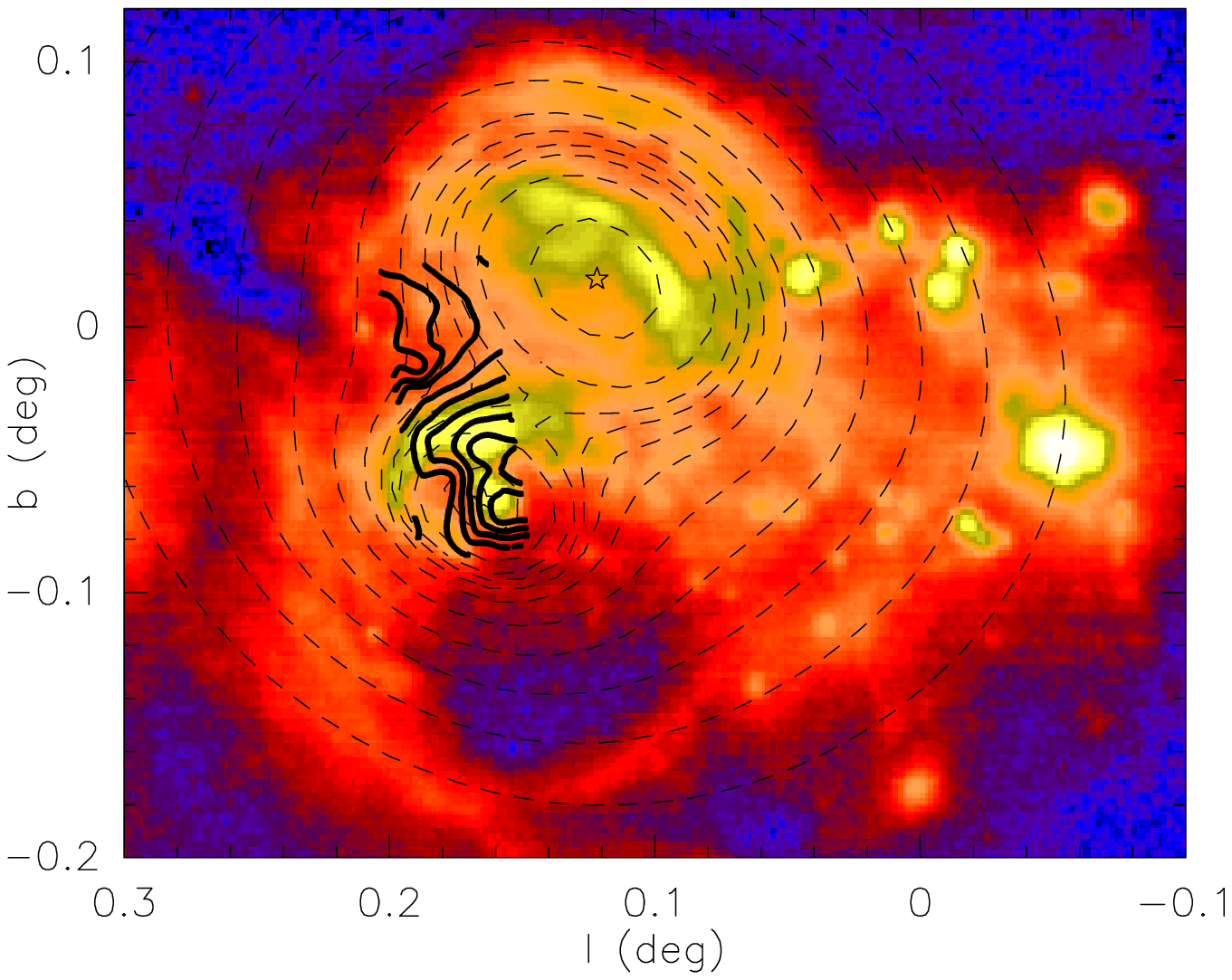}
}
\caption{Dust and ionized gas in the Radio Arc region (see text).}
\label{figmap}
\end{figure}

Here, we present the first large scale study of the ionization structure
in the Radio Arc region.  
We have analyzed fine structure lines observations made by the
Infrared Space Observatory (ISO) toward the sources
shown in the left panel of  Fig. \ref{figmap} with filled squares.

The right panel of Fig. 1  shows, with thick contours, 
a map of the {N\,\sc iii} 57~$\mu$m to {N\,\sc ii} 122~$\mu$m lines
ratio (hereafter {N\,\sc iii}/{N\,\sc ii} ratio).
The map shows two clear gradients, one  pointing to the Quintuplet
(in the southern  part of the map) and
other pointing to the   Arches cluster (indicated by a star).

From the {O\,\sc iii} 52/88 $\mu$m line ratio we derive an electron density
($n_e$) of $\sim 200$ cm$^{-3}$ for all the sources.
However, photoionization modeling shows that the interstellar medium (ISM) in
this region cannot have an average density of $\sim 200$ cm$^{-3}$ in order
to explain the size of the ionized region, which is larger than 30 pc.
Instead, the ISM should be rather inhomogeneous.
Thus, we have considered the sources as independent clouds with a density
of 200~cm$^{-3}$ located at a distance to the clusters equal to their
projected distance.
We have found that the trend of the {N\,\sc iii}/{N\,\sc ii} ratio
with the distance from the Quintuplet cluster is consistent with the flux of
Lyman continuum photons ($Q$) estimated from the stellar content of the cluster
($Q\sim 10^{50.9}$~s$^{-1}$) and  effective temperatures ($T_{eff}$) of
$\sim 33000$~K (see RF01).
For the sources that are clearly influenced by the Arches cluster,
the trend of the {N\,\sc iii}/{N\,\sc ii} 
ratio with the distance to the cluster
is also consistent with the cluster parameters ($Q\sim 10^{51.4}$~s$^{-1}$)
for $T_{eff}\sim33000$~K (RF01).

Since those $T_{eff}$  are rather similar for both clusters,
to estimate  the combined effect of the clusters we have used a very
simple model in terms of a total ionization parameter ($U$) defined as the sum
of the ionization parameters due to the Quintuplet  and
the Arches clusters. Thus, for a cloud located at distances
$D_Q$ and $D_A$ from the Quintuplet and the Arches clusters, respectively,
$U$ will be given by:
$ U= \frac{Q_Q}{4 \pi D_Q^2 n_e c} + \frac{Q_A}{4 \pi D_A^2 n_e c} $,
where $c$ is the velocity of light.
The right panel of Fig. 1 shows as dashed lines some contours of equal $U$ assuming
$n_e=10^{2.2}$~cm$^{-3}$, $Q_Q=10^{50.9}$~s$^{-1}$ and 
$Q_A=10^{51.4}$~s$^{-1}$.
The agreement of the iso-$U$ curves with the {N\,\sc iii}/{N\,\sc ii} map
is very good taking into account the simplicity of the model.
Furthermore, this simple model also reproduces the 
distribution of warm dust as observed by MSX (the 
background image  of Fig. 1 is the $\sim 20~\mu$m MSX image).

We conclude that the large scale  ionization and the heating of the dust
in the Radio Arc region are 
dominated by  the combined effect of the Quintuplet and the  Arches clusters.
They ionize a large region of more than $30\times30$~pc$^2$.
Any other possible ionization
mechanism as interaction with  magnetic fields
or more hot stars must play a minor role.






\end{article}
\end{document}